\documentclass[twoside,11pt]{article}

\usepackage{jmlr2e}

\usepackage{url}
\PassOptionsToPackage{hyphens}{url}\usepackage{hyperref}
\usepackage{cite}

\usepackage{xcolor}
\usepackage{array}
\usepackage{algorithmic}
\usepackage{tabularx}
\usepackage{amssymb}
\usepackage{pifont}
\usepackage{float}
\usepackage{booktabs}

\jmlrheading{166}{2022}{1-24}{6/22}{12/22}{Koutini et al.}

\ShortHeadings{Learning Audio Representations With PaSST}{Koutini et al.}
\firstpageno{1}

\begin{document}

\title{Learning General Audio Representations \\With Large-Scale Training of Patchout Audio Transformers}

\author{\name Khaled Koutini$^{1,2}$ \\
        \name Shahed Masoudian $^{1,2}$ \\
        \name Florian Schmid  $^{1,2}$ \\
        \name Hamid Eghbal-zadeh$^{1,2}$ \\
        \name Jan Schlüter$^{1}$ \\
        \name Gerhard Widmer$^{1,2}$ \\
      \addr Institute of Computational Perception$^1$ \& LIT AI Lab$^2$ 
      \\
      Johannes Kepler University\\
      Linz, Austria
}

\editor{Joseph Turian, Björn W. Schuller, Dorien Herremans, Katrin  Kirchoff,  Paola Garcia Perera, Philippe  Esling}

\maketitle

\begin{abstract}
The success of supervised deep learning methods is largely due to their ability to learn relevant features from raw data. 
Deep Neural Networks (DNNs) trained on large-scale datasets are capable of capturing a diverse set of features, and learning a representation that can generalize onto unseen tasks and datasets that are from the same domain.
Hence, these models can be used as powerful feature extractors, in combination with shallower models as classifiers, for smaller tasks and datasets where the amount of training data is insufficient for learning an end-to-end model from scratch.
During the past years, Convolutional Neural Networks (CNNs) have largely been the method of choice for audio processing. However, recently attention-based transformer models have demonstrated great potential in supervised settings, outperforming CNNs.
In this work, we investigate the use of audio transformers trained on large-scale datasets to learn general-purpose representations.
We study how the different setups in these audio transformers affect the quality of their embeddings. We experiment with the models' time resolution, extracted embedding level, and receptive fields in order to see how they affect performance on a variety of tasks and datasets, following the HEAR 2021 NeurIPS challenge evaluation setup.
Our results show that representations extracted by audio transformers outperform CNN representations. 
Furthermore, we will show that transformers trained on Audioset can be extremely effective representation extractors for a wide range of downstream tasks.
\end{abstract}


\section{Introduction}

The high cost and scarcity of high-quality labeled data is a significant impediment to applying deep learning methods to audio classification and tagging tasks. 
One popular alternative is to use models pre-trained on large-scale and broad datasets~\citep{audioset2017Gemmeke,youtub8m16}  to obtain high-quality and relevant representations from raw waveforms, which can improve performance on downstream tasks.
Audioset~\citep{audioset2017Gemmeke} is a dataset of annotated clips from Youtube videos. The dataset contains approximately 2 million audio clips of 10 seconds each. Each clip is assigned tags representing the acoustic events occurring in the clip. There are 527 possible classes that are taken from the Audioset ontology~\citep{audioset2017Gemmeke}. The ontology covers a wide variety of events that include human and animal sounds, music, sounds of nature and things, and environments and backgrounds. Given the size and diversity of the dataset, it has been widely used to pre-train acoustic models with the goal of either fine-tuning these models on smaller datasets or  extracting general features and representations out of raw audio waveforms~\citep{KongCIWWP20panns,yammnet,cramerWSB19openl3,koutini21passt}.

The most common approaches for extracting audio presentation from raw waveforms has been to use pre-trained architectures based on Convolutional Neural Networks (CNNs)~\citep{hersheyCNNArchitecturesLargescale2017,cramerWSB19openl3,SchneiderBCA19wave2vec_orig}. This is primarily due to the success of CNNs in various audio processing tasks and applications~\citep{KongCIWWP20panns,hersheyCNNArchitecturesLargescale2017,koutini21journal}. VGGish~\citep{hersheyCNNArchitecturesLargescale2017} is an example of a large-scale pre-trained CNN  that is used to obtain features on which a task-specific classifier is trained~\citep{humphrey2018openmic}.
\citet{ArandjelovicZ17L3net} propose a self-supervised multi-modal approach (L$^3$-Net) to learn audio representations without the need for labels. 
\citet{cramerWSB19openl3} use AudioSet
to train L$^3$-Net, resulting in representations that outperform the VGGish model on a variety of tasks.
\citet{KongCIWWP20panns} introduced an improved CNN architecture that significantly improves on the VGGish
model on Audioset.
They also test transfer learning from models trained on Audioset to other tasks. They show that either fine-tuning or using the pre-trained CNN as a feature extractor can both help on some downstream tasks, such as acoustic scene classification and music genre classification. 
 

\textit{Transformer models}~\citep{vaswani2017attention} have shown to be quite effective and have improved upon CNNs in many domains. Transformers enable the learning of relationships between distinct elements in sequences, independent of their placement or distances. Transformers have delivered state-of-the-art results in a variety of natural language processing applications~\citep{devlinCLT19bert}. The architecture was ported to computer vision~\citep{dosovitskiyB0WZ21VIT} by extracting patches from the input images and augmenting each patch using a learnable positional encoding. The resulting patches create a sequence that can be processed by a Transformer model.

Vision transformer models achieve state-of-the-art performance on different image classification tasks\citep{dosovitskiyB0WZ21VIT,TouvronCDMSJ21deit} as well as audio tagging and classification tasks~\citep{gong21ast,koutini21passt}.  However, they have a number of disadvantages, including high training complexity and cost, the need for a large amount of data, and the difficulty of adapting the learned positional encodings for variable-sized input.
In order to overcome these problems, a large private dataset was used in the case of Vision Transformer (Vit)~\citep{dosovitskiyB0WZ21VIT}. The authors of Data-efficient Image Transformers (DeiT)~\citep{TouvronCDMSJ21deit} used a wide range of data augmentation methods and knowledge distillation from CNNs. The authors of Audio Spectrogram Transformer (AST)~\citep{gong21ast} leveraged the models' pre-training on vision tasks to train on spectrograms. 
Patchout Audio Transformer (PaSST)~\citep{koutini21passt} addresses the problem of training complexity by significantly lowering the compute and memory requirements for spectrogram training with \textit{Patchout}. Patchout is also a regularization method that improves the models' generalization. PaSST separates time and frequency positional encodings, allowing for simple inference on shorter audio clips without the need to re-train or interpolate positional encodings.


In this paper, we investigate the extraction of general-purpose representations from pre-trained PaSST. We evaluate the extracted representations using HEAR-eval~\citep{turian2022hear} on a wide range of audio tasks.  We investigate various representation levels, spectrogram time resolutions, and input receptive fields. We compare the quality of our representation to that of widely used models. The study results will demonstrate that large-scale supervised training of transformers can yield powerful feature extractors  that outperform CNNs trained on the same dataset and are competitive with task-specific models.

\section{Patchout Fast Spectrogram Transformer (PaSST)}
In this section, we will go over the model architecture as well as the training approach. The Patchout faSt Spectrogram Transformer (PaSST)~\citep{koutini21passt}, which is based on the Vision Transformer architecture~\citep{dosovitskiyB0WZ21VIT}, is used. Applying Patchout~\citep{koutini21passt} during model training significantly reduces model training time and acts as a regularizer to improve model generalization. Furthermore, disentangled time and frequency positional encodings in PaSST are critical for predicting on shorter audio clips and restricting the model's receptive field, which is required to produce timestamp embeddings.

\subsection{Vision Transformers}
\begin{figure}[t]
\centering
{\includegraphics[width=0.7\textwidth]{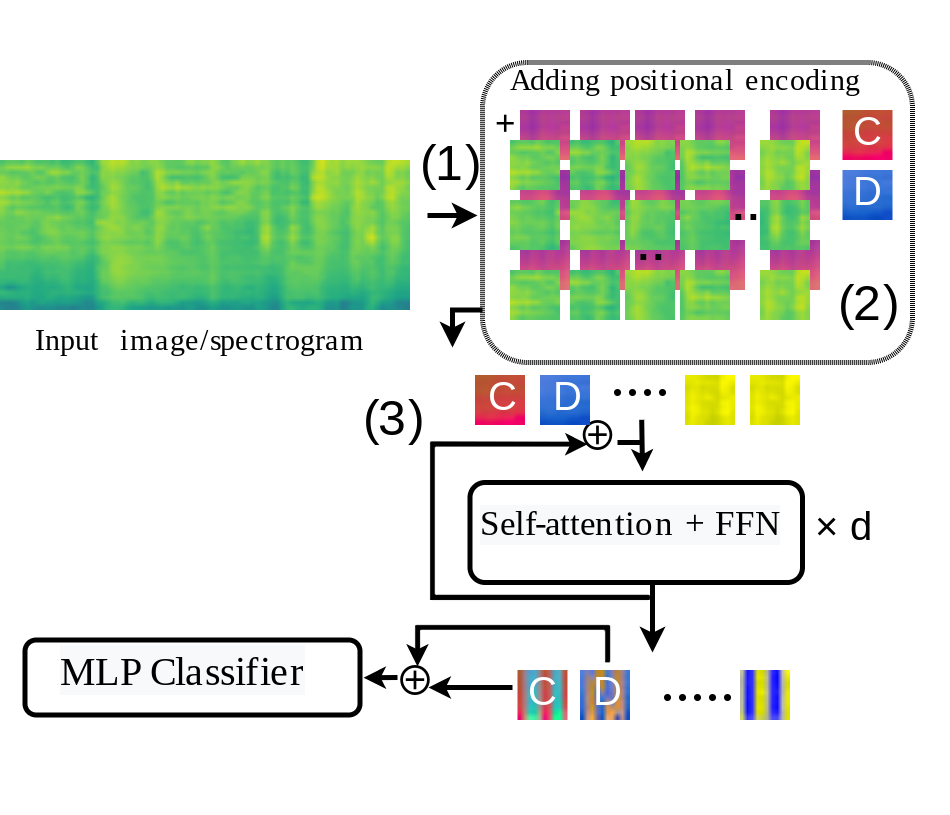}}
\caption{The Architecture of ViT~\citep{dosovitskiyB0WZ21VIT}, DeiT~\citep{TouvronCDMSJ21deit} and AST~\citep{gong21ast}. C,D are the classification and distillation tokens. }
\label{fig:vit}
\end{figure}

The Vision Transformer (ViT)~\citep{dosovitskiyB0WZ21VIT} extracts small patches from an input image and projects them linearly onto a sequence of embeddings (Step 1 in Figure~\ref{fig:vit}). Each element's position in the sequence is then provided in the form of a bias, which is added to each embedding (Step 2 in Figure~\ref{fig:vit}). Each of these positional encodings (biases) is a trainable parameter that aids in carrying information about the patch's position in the original image~\citep{dosovitskiyB0WZ21VIT}. The embedding sequence is supplemented with a special classification embedding (classification token C in Figure~\ref{fig:vit}). Following the self-attention layers, the classification token is linked to a classifier. 
Another special embedding for distillation (distillation token D in Figure~\ref{fig:vit}) is added in Data-efficient image Transformers (DeiT)~\citep{TouvronCDMSJ21deit}. DeiT has a similar architecture to ViT, but uses CNN distillation and data augmentation to reduce the amount of training data needed to train the transformer models. Audio Spectrogram Transformer (AST)~\citep{gong21ast} uses the same architecture applied to audio spectrograms.  AST leverages pre-training on vision datasets to achieve good performance on Audioset~\citep{audioset2017Gemmeke}.

\begin{figure}[tbh]
\centering
{\includegraphics[width=0.7\textwidth]{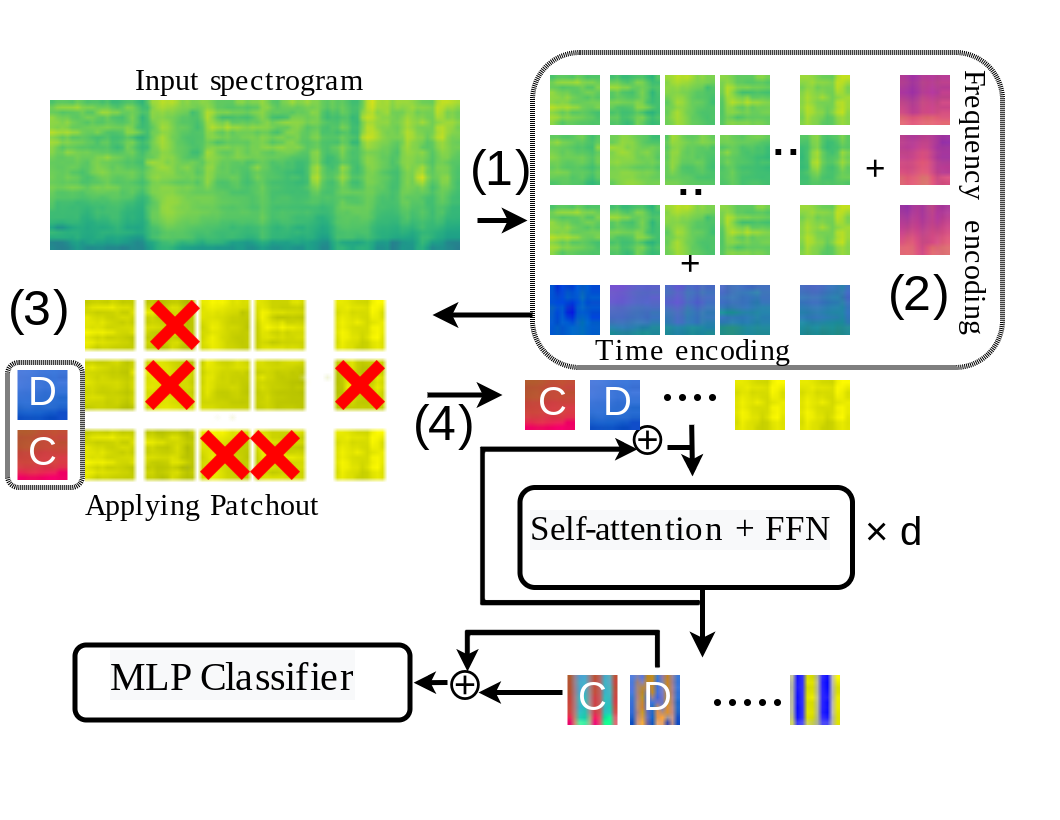}}
\caption{The Architecture of PaSST~\citep{koutini21passt}. The distinction between Figures~\ref{fig:vit} and~\ref{fig:passt} is in disentangling the positional encodings into time and frequency positional encodings. In addition, patchout is used during training. }
\label{fig:passt}
\end{figure}

PaSST~\citep{koutini21passt}\footnote{Source code available: \url{https://github.com/kkoutini/PaSST}} differs from previous models in two ways  (Figure~\ref{fig:passt}): first, it disentangles positional encodings into time and frequency positional encodings. Second, during training, a regularization strategy named \textit{Patchout} is used.
The following sections will go into these in detail. In summary, PaSST operates in the manner depicted in Figure~\ref{fig:passt}: (1) Extract patches from the spectrogram (2) Augment these patches with time and frequency encondings (3) Use Patchout, and then (4) run the resulting sequence through the transformer.

\subsection{Patchout}
\label{sec:patchout}
Patchout~\citep{koutini21passt} works on the basic principle of dropping parts of the transformer's input sequence during training, training the transformer to perform classification using an incomplete sequence.
Reducing the sequence length both regularizes the model and significantly reduces the complexity of the training process. During inference, the entire input sequence is presented to the transformer, similar to DropOut~\citep{JMLR:v15:srivastava14a:dropout}. We distinguish the following types of Patchout: 

\textbf{Unstructured Patchout:} is the most basic form, in which the patches are chosen at random regardless of their position.

\textbf{Structured Patchout:}  We select some frequency bins / time frames at random and remove an entire column or row of extracted patches, similar in spirit to SpecAugment~\citep{ParkCZCZCL19Specaugment}. In the remainder of this work, we use only \emph{Structured Patchout} since \citet{koutini21passt} showed that it performs better on Audioset.

\begin{figure}[t]
\centering
{\includegraphics[width=0.7\textwidth]{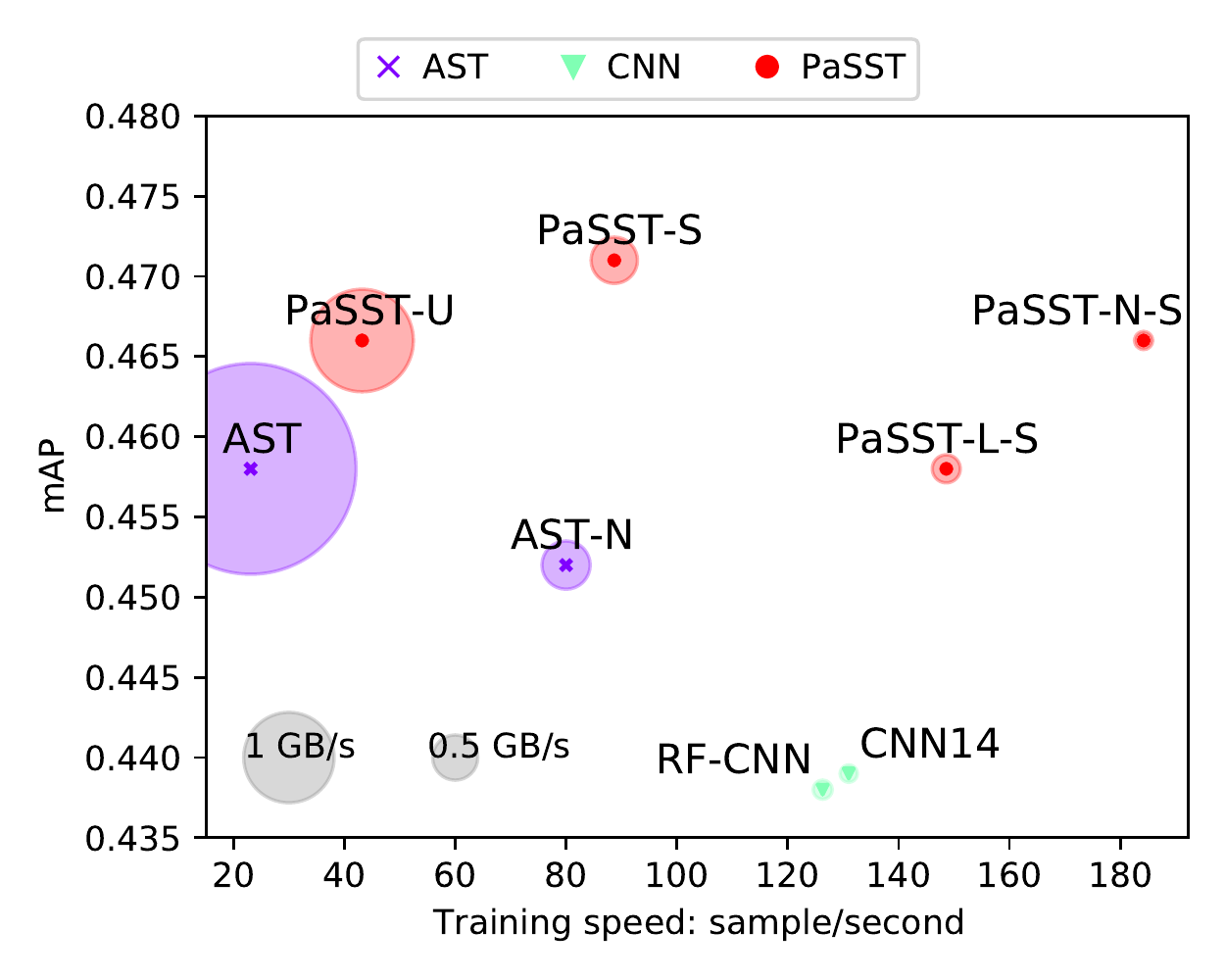}}
\caption{ Prediction performance on Audioset (mean average precision) vs.~training speed (samples per second). The radius of the circle indicates the required GPU memory (in gigabytes) per training sample. The figure shows PaSST variants~\citep{koutini21passt} with \emph{Structured Patchout} (denoted by \textbf{-S}) and \emph{Unstructured Patchout} (denoted by \textbf{-U}) compared with AST~\citep{gong21ast}, CNN14~\citep{KongCIWWP20panns} and RF-CNNs~\citep{Koutinitrrfcnns2019}. The PaSST variants marked with -L and -N are low-complexity and no-patch-overlap variants, respectively. PaSST-S achieves state-of-the-art Audioset performance in less than two days on a single consumer GPU. Source:~\citep[Figure 1]{koutini21passt}. }
\label{fig:passtspeed}
\end{figure}

Figure~\ref{fig:passtspeed} shows that applying Patchout significantly reduces the training complexity while improving generalization. The figure shows that with structured Patchout, PaSST-S achieves higher performance on Audioset with 4 times the training speed of AST~\citep{gong21ast} and a fraction of the required memory. Furthermore, PaSST-N-S (with structured patchout and no patch overlap) outperforms CNNs with similar memory requirements and faster training.

In this paper, we will use the  \emph{PaSST-S} variant, which has the best performance on Audioset. Section~\ref{sec:extract} discusses how we extract the embeddings of this model. 

\subsubsection{Computational complexity}
\label{sec:passt:complex}
Transformers primary building blocks are attention layers~\citep{vaswani2017attention}.
Attention layers compute a distance between each pair of items in the input sequence, and these distances form the attention weights, resulting in a quadratic complexity with respect to the input sequence length, which in turn is a function of the length of the audio clips, the spectrogram resolution in time or frequency, and the patches overlap.
 As the sequence length $n$ grows, the computation complexity (and memory requirements) for the attention layers grows quadratically  $\mathcal{O}(n^2)$ ~\citep[Section 2.1]{koutini21passt}.
In short, reducing the length of the input sequences to self-attention layers would have a significant impact on the computational complexity of these models; Patchout addresses this issue.

\subsubsection{Regularization Effect}
 
Previous work has shown that, in the case of CNNs, restricting the receptive field on audio spectrograms during training improves their generalization on a variety of audio tagging and classification tasks~\citep{koutini21journal}. Patchout has a similar effect on transformers by removing parts of the input spectrograms at random during training.
While in CNNs the receptive field can be easily controlled via depth and kernel sizes~\citep{koutini21journal}, transformers by design have a large receptive field and exploit distant dependencies in their input. That allows them to learn complex rules from their input, but can also lead to over-fitting, which is one of the reasons why transformers require a large amount of data and augmentations. Patchout ameliorates this problem by forcing the transformer to learn from incomplete and augmented spectrograms. Structured Patchout, specifically, is also similar to SpecAugment~\citep{ParkCZCZCL19Specaugment}, a data augmentation technique specific to spectrograms that helps in various audio tasks. \citet{koutini21passt} showed that Patchout leads to improvements in different datasets.

\subsection{Disentangling Positional Encoding}
\label{sec:passt:disent}
The second distinction between PaSST~\citep{koutini21passt} and vision transformers~\citep{dosovitskiyB0WZ21VIT,TouvronCDMSJ21deit,gong21ast} is that in PaSST, the positional encoding is disentangled into frequency and time encoding (see Figures~\ref{fig:vit} and ~\ref{fig:passt}). This simplifies predicting on audio clips of varying lengths, due to the fact that the learned frequency encoding does not need to be changed and is unaffected by the length of the audio clip. All that is needed is is to change the time encoding. Cropping the time encoding- to match the length of the audio clip- is a simple yet effective way for inference. This, however, does not address the issue of inference on audio clips that are longer than the ones on which the model was trained (10 seconds in Audioset), because the time encoding is only learned to cover 10 seconds of audio during training. As explained in Section~\ref{sec:extract}, we address this by taking the mean of the embeddings extracted from overlapping 10-second windows. We leave learning time encodings for longer audio clips using improved training algorithms for future work.

\section{General Evaluation of Audio Representations}
\label{sec:hear:eval}
This section explains the diverse audio tasks and the setup used to evaluate the representations extracted from the pre-trained PaSST models. We use the same setup and tasks used in the Holistic Evaluation of Audio Representations (HEAR) NeurIPS 2021 challenge.

\subsection{Evaluation Tasks}
\label{sec:hear:eval:Tasks}

 The goal is to evaluate the models based on the quality of their generated representations across a wide range of tasks and scenarios. The tasks involve both short and long time spans and cover a wide range of audio domains, including speech, environmental sound, and music. 
 
 HEAR consists of nineteen diverse tasks\footnote{The detailed description of the tasks and the datasets can be found on the HEAR benchmark website: \url{https://hearbenchmark.com/hear-tasks.html}}. \citet{turian2022hear} provide a detailed description of the tasks and the setup of each task for HEAR evaluation. 
 We divided the tasks into three categories based on the audio domain and summarize the tasks and their datasets below: 
\begin{itemize}
    \item \textbf{Speech:} Contains all the tasks that consist of verbal articulations by humans.
    \begin{itemize}
        \item  \textbf{Speech Commands 5hr:} Spoken command recognition and classification, described in~\citep{Warden18Speech}. 
        \item  \textbf{Speech Commands Full:}  Spoken command recognition and classification using 27 hours of data. 
        \item \textbf{CREMA-D:} The goal is to classify one of six different emotions (anger, disgust, fear, happiness, neutral, or sad) from a spoken sentence~\citep{CaoCKGNV14}. The dataset contains $7,438$ 5-second audio clips.
        \item \textbf{LibriCount:} The goal of this task is to count the number of speakers in an audio clip~\citep{StoterCEH18}. The dataset contains $5,720$ 5-second audio clips.
        \item \textbf{VoxLingua107 Top10:} The task is to classify the language spoken in an audio clip into one of ten different languages~\citep{ValkA21}. There are $972$ clips totaling about five hours of audio in the dataset.
    \end{itemize}

\item \textbf{Music:} Contains all tasks consisting of instrumental sounds:
    \begin{itemize}
        \item  \textbf{NSynth Pitch 5hr:}  Instrumental sound classification from the NSynth Dataset~\citep{nsynth2017} into one of 88 pitches.
        \item  \textbf{NSynth Pitch 50hr:}  Similar to \textbf{NSynth Pitch 5hr} but with more training data.
        \item  \textbf{Beijing Opera Percussion:} The goal is to classify four main categories of percussion instruments~\citep{TianSSS14}. The dataset contains $236$ 5-second audio clips.
        \item  \textbf{GTZAN Genre:} The task is to classify a 30-second audio track into one of ten possible genres~\citep{TzanetakisC02}. The dataset contains $1,000$ tracks.
        \item \textbf{MAESTRO 5hr:} The task is to perform music transcription using the MAESTRO dataset~\citep{HawthorneSRSHDE19maestro}.
        \item \textbf{Mridangam Stroke:} The goal is to classify 10 different strokes using the Mridangam Stroke Dataset~\citep{AnantapadmanabhanBM13}. The dataset contains $6,977$ audio clips.
        \item \textbf{Mridangam Tonic:} The goal is to classify the 6 different tonics of the Mridangam Stroke Dataset.
        
    \end{itemize}

\item \textbf{General:} Contains mostly environmental sounds and acoustic scenes, but also others that cannot be clearly put into either \textit{Speech} or \textit{Music} (like GTZAN Music Speech):
    \begin{itemize}
        \item  \textbf{DCASE 2016 Task 2:} Office sound event detection task, adapted from the DCASE 2016 Challenge~\citep{Lafay2017}. The dataset consists of 72 audio clips, each lasting 2 minutes.
         \item  \textbf{Beehive States:} The goal is to identify the absence of the queen in a beehive~\citep{NolascoTCOBB19}. This dataset contains 930 clips that are approximately 10 minutes long.
         \item  \textbf{ESC-50:} The task is environmental sound classification ~\citep{piczak2015dataset}. The dataset contains $2,000$ 5-second audio clips, each of which is classified into one of fifty categories.
         \item  \textbf{FSD50K:} The task is a general-purpose multi-label sound event detection task~\citep{fonsecaFPFS22FSD50K}. The events are classified into 200 classes, including  environmental sounds, speech, and music. The dataset contains $51,000$ audio clips of variable length. 
         \item \textbf{Gunshot Triangulation:} The goal is to identify the location of the microphone recording a gunshot. The gunshot is recorded using microphones at different distances from the shooter~\citep{cooper10gunshots}.   The dataset is small in size, with 22 shots from 7 different firearms totaling 88 audio clips of 1.5 seconds each.
         \item  \textbf{GTZAN Music/Speech:} The task is to classify a 30-second audio track into music or speech. The dataset consists of $120$ tracks~\citep{TzanetakisC02}. 
         \item \textbf{Vocal Imitations:} The task is to retrieve a sound by using its vocal imitation~\citep{dcase/KimGPD18}. The dataset contains $5,601$ audio clips, each a vocal imitation of one of $302$ reference sounds.
    \end{itemize}
\end{itemize}

\subsection{Evaluation Setup}
\label{sec:hear:eval:setup}

The HEAR-eval~\citep{turian2022hear} tool was used to evaluate the representations extracted from PaSST models.  We also compare the findings to the official HEAR 2021 challenge results, which were obtained using the same evaluation tool. 

The HEAR-eval tool uses the representations extracted from the evaluated models to train a shallow downstream multi-layer perceptron (MLP) classifier.
In other words, the evaluated models are only employed in the process of extracting features from the audio inputs. These features are then frozen and used to train a task-specific MLP model with the task labels as targets.
Afterward, the trained MLP is tested using the extracted representations from the task test set. I refer the reader to ~\citep{turian2022hear} for more details about the HEAR evaluation setup.

The different tasks have different evaluation metrics, such as accuracy and mean average precision (mAP). Therefore, we normalize the evaluation score in order to group and compare the results of individual tasks into the three main categories, speech, music, and general. The test scores of each task are normalized across the evaluated models so that the maximum test score achieved by a model in the official HEAR 2021 challenge corresponds to $1$. As a result, after normalization, the performance of a model on a task is expressed as a percentage of the best-performing system in the official challenge.

The HEAR evaluation tool can test two kinds of representations, depending on the task requirements:
\begin{description}
    \item[Timestamp Embedding:] The timestamp embeddings are localized embeddings at regular intervals, representing the information contained in short periods of a longer audio clip. Timestamp embeddings are useful for the tasks that rely on time-local information, such as sound event localization and music transcription. There are only two HEAR tasks that depend on the timestamp embeddings: \emph{DCASE 2016 Task 2} and \emph{MAESTRO 5hr}.
    \item[Scene Embedding:] A scene embedding is a single representation that summarizes all the information contained in an audio clip, regardless of its length. The scene embeddings are useful for the tasks that aim to classify the input audio or tag the whole audio clip by its contents. The majority of HEAR tasks use scene embeddings (all the tasks except the two mentioned above).
    
\end{description}

\section{Extracting Representations from PaSST}
\label{sec:extract}
This section describes how we extract audio representations from PaSST models  that have been pre-trained on Audioset. For pre-training, we chose a supervised multi-class multi-label prediction task. The task is to predict tags for 10-second audio clips that represent the presence or absence of 527 different classes. The models are trained on the balanced and unbalanced training subsets of Audioset. The detailed training setup are presented in Appendix~\ref{app:setup}.

Section~\ref{sec:extract:embed:level} explains the three levels of representation that can be extracted from the trained model. Section~\ref{sec:extract:temp:resolution} discusses how to change the transformer model's temporal resolution and how this affects its performance and complexity. Sections ~\ref{sec:extract:scene} and ~\ref{sec:extract:timestamp} describe how to extract localized and general embedding for the HEAR challenge~\citep{turian2022hear}.

\subsection{Representation Extraction Levels}
\label{sec:extract:embed:level}

\subsubsection{High-Level representations: Logits}  
Logits are the output of the model's final layer (the MLP classifier in Figure~\ref{fig:passt}) and contain information about the classes used in supervised training.
Given the variety of classes in Audioset, we anticipate that the information in the models' logits will be useful for a variety of the benchmark tasks.

\subsubsection{Mid-Level representations: Features}
The processed classification and distillation tokens from the self-attention layers are fed into the final MLP classifier (shown in Figure~\ref{fig:passt}).  As a result, we expect these transformed tokens to learn to convey the information required to predict the classes of the supervised task from the input audio.

\subsubsection{Low-Level representations: Mel features}

We look into the inclusion of low-level features. Low-level features are mel scaled spectrograms that we feed into the transformer model. We add mel-spectrograms from a small window around the timestamp in timestamp embedding. 
We hypothesize that mel-spectrograms can improve the performance in novel tasks to PaSST, where high-level representations may lack relevant information.

\subsection{Temporal Resolution}
\label{sec:extract:temp:resolution}
PaSST's base model extracts $16 \times 16$ patches from the input~\citep{koutini21passt}. As a result, the input spectrogram should have at least  $16$ time frames in optimal settings. This, however, is dependent on the audio pre-processing setup and Short-time Fourier transform hyper-parameters (STFT) (Details in Appendix~\ref{app:setup}).
More specifically, the hop length of the STFT window determines the audio length covered by the $16$ time frames and thus the minimum audio length that PaSST can optimally process.

STFT window hop lengths of  $10$ ms (PaSST default), 5 ms, and 3 ms are investigated.
However, increasing the temporal resolution has the unintended consequence of increasing computational complexity because the model must process higher dimensional input representing the same amount of time.
Particularly in the case of a transformer, where a higher-resolution spectrogram results in longer sequences for attention layers, resulting in a quadratic increase in compute and memory complexity (see Section~\ref{sec:passt:complex}). Patchout can help make training on higher resolution spectrograms feasible by providing a $9.9 \times $ speed increase (Table~\ref{tab:time:resolution}).

\subsection{Extracting Scene Embedding from PaSST}
\label{sec:extract:scene}
The majority of HEAR's tasks rely on scene embedding~\citep{turian2022hear}.
Scene embedding is a single representation of an audio clip that summarizes its entirety. This is useful for tasks such as classification and tagging for the content of an entire audio clip. 
PaSST was trained on general audio tagging tasks using Audioset. As a result, we anticipate that the extracted representations will summarize the information relevant to Audioset classes present in the model's input. 

We extract low-, mid-, and high-level representations ( Section~\ref{sec:extract:embed:level}) from the pre-trained PaSST models on different time resolutions (Section~\ref{sec:extract:temp:resolution}).
Because the model was trained on 10-second audio clips, and because of the disentangled time and frequency positional encoding, PaSST can process 10-second or shorter audio clips (see Section~\ref{sec:passt:disent}).
For longer inputs, we simply take the mean of the embeddings of 10-second windows with overlap.


\subsection{Extracting Timestamp Embedding from PaSST}
\label{sec:extract:timestamp}

Timestamp embedding represents information in the input audio at regular intervals. This can be useful for event detection, localization, and predicting onsets and offsets. 

The receptive field of the final layers of a fully convolutional CNN on the input is the accumulative receptive field of all the preceding layers~\citep{Koutini2019Receptive}. As a result, mapping the generated representations in such networks to the corresponding audio chunks is trivial. Transformers, on the other hand, dynamically route information through the attention layers from all over the input. Therefore, no such trivial mappings can be computed.
We choose to chunk the raw audio waveforms to windows with a 50 ms hop (as recommended in~\citet{turian2022hear}). These windows represent the receptive field of the transformer around the center of each time interval.

In HEAR~\citep{turian2022hear}, DCASE 2016 Task 2 and MAESTRO rely on timestamp embedding. We investigate different levels of representation as embeddings. We test models with various receptive fields as well as a concatenation of representations extracted from two receptive fields.

\section{Results}
\label{sec:results}
The HEAR-eval~\citep{turian2022hear} tool was used to generate the results for all PaSST models. For all other models presented in this section, the official results of the HEAR 2021 challenge are used, which were produced using the same tool.
Section~\ref{sec:hear:eval} provides details about the evaluation setup used to obtain the following results. In short, we divided the tasks into three categories based on the audio domain (see Table~\ref{table:task_catgory}):
\begin{itemize}
    \item \textbf{Speech:} Contains all the datasets that consist of verbal articulation of humans.
    \item \textbf{Music:} Contains all datasets consisting of instrumental sounds. 
    \item \textbf{General:} Contains mostly environmental sounds and acoustic scenes, but also others that cannot be clearly put into either \textit{Speech} or \textit{Music} (like GTZAN Music Speech). 
\end{itemize}

\begin{table}[tbh!]
\centering
\caption{ HEAR 2021 tasks \citep{turian2022hear} and the categories we assign them to.}
\label{table:task_catgory}
\begin{tabular}{ |m{12em}| c| c| c| }
 \hline
 \textbf{Task} &  \textbf{Music} & \textbf{Speech} &  \textbf{General} \\
 \hline
 Beehive States &  &  & x  \\ 
 Beijing Opera Percussion & x &  &   \\ 
 CREMA-D &  & x &  \\ 
 DCASE 2016 Task 2 &  &  & x \\ 
 ESC-50 &  &  & x \\  
 FSD50K &  &  & x \\  
 GTZAN Genre & x &  &  \\
 GTZAN Music Speech &  &  & x \\
 Gunshot Triangulation &  &  & x \\
 LibriCount &  & x &  \\  
 MAESTRO 5h & x &  &  \\  
 Mridingham Stroke & x &  &  \\  
 Mridingham Tonic & x &  &  \\  
 NSynth Pitch 5h & x &  &  \\  
 NSynth Pitch 50h & x &  &  \\ 
 Speech Commands 5h &  & x &  \\
 Speech Commands full &  & x &  \\
 Vocal Imitations &  &  & x \\ 
 VoxLingua107 Top 10 &  & x &  \\ 
 \hline
\end{tabular}
\end{table}

For each of the 19 tasks, different metrics and ranges are used to evaluate them. As a consequence, whenever tasks in the following plots are grouped into categories, we normalize the test score so that the maximum test score achieved by a model in the official HEAR 2021 challenge~\citep{turian2022hear} corresponds to $1$. 
In the following, we analyze the use of low-level features and different receptive field sizes for the two timestamp tasks: \textit{DCASE 2016 Task 2} and \textit{MAESTRO}. We further compare different models on the three outlined task categories, the use of different levels of representation, and different temporal resolutions. This comparison is conducted on both timestamp and scene embedding tasks. 

The abbreviations used to describe the models are as follows: \textit{L, M} and \textit{H} indicate the use of low-, medium- and high-level features, respectively (see Section~\ref{sec:extract:embed:level}); \textit{hop} describes the hop size of the STFT window; and \textit{2RF} indicates the concatenation of two embeddings with two different receptive fields. The default values used in the baseline model of PaSST~\citep{koutini21passt} (challenge name: \textit{CP-JKU PaSST base}) are features \textit{M + H}, a hop size of \textit{10 ms} and a receptive field size of \textit{160 ms}. It is below denoted as \textit{PaSST M+H hop=10 ms}.

\subsection{Timestamp Embedding Tasks}

\begin{figure}[bth!]
\centering
{\includegraphics[width=1.0\textwidth]{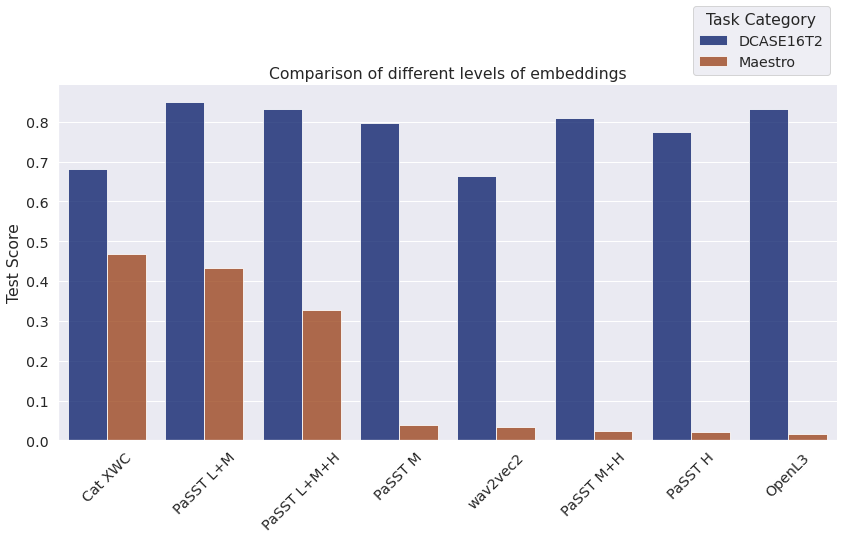}}
\caption{Comparison of different levels of extracted representations for PaSST models with \textit{hop=10 ms} and receptive field size of \textit{160 ms}.}
\label{fig:dm_base}
\end{figure}

Two tasks in HEAR 2021, \textit{DCASE 2016 Task 2} and  \textit{MAESTRO}, rely on timestamp embeddings. We compare changes in our setup that effect timestamp embeddings with two single models,  OpenL3~\citep{cramerWSB19openl3} and Wav2Vec2~\citep{baevski2020wav2vec}, as well as the best performing submission to HEAR 2021 in the \textit{MAESTRO} task, named \textit{Cat XWC}, as a reference.
OpenL3 is trained in a self-supervised way using the audio and videos of Audioset.
\textit{Cat XWC} is a concatenation of three diverse models: (1) \textit{HuBERT} \citep{hsu2021hubert} which is a self-supervised approach for speech representation learning, (2) \textit{wav2vec2}\citep{baevski2020wav2vec} is another model pre-trained on speech tasks (3) \textit{CREPE} \citep{kim2018crepe} which is a specialized pitch estimation model.

\subsubsection{Adding Low-Level Features }

Figure \ref{fig:dm_base} shows that including low-level embeddings (L+M and L+M+H) outperforms approaches with other combinations on the \textit{MAESTRO} task while maintaining a high performance on the \textit{DCASE} task. Models that do not include the low-level features fail to achieve a high score on the \textit{MAESTRO} task, which means that the mid- and high-level representations extracted by \textit{PaSST} do not convey relevant information for the music-related \textit{MAESTRO} task. 

\subsubsection{Receptive Field }
\begin{figure}[tbh!]
\centering
{\includegraphics[width=1.0\textwidth]{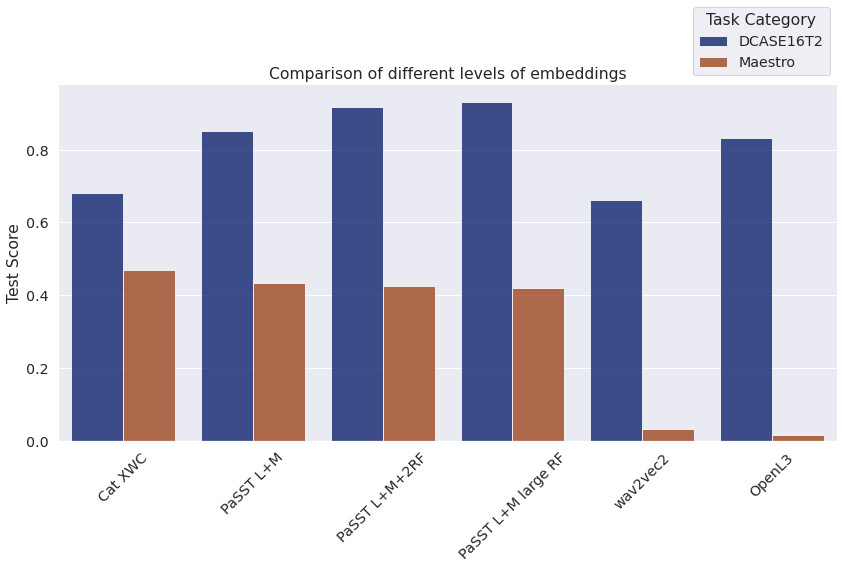}}
\caption{Comparison of different receptive field (RF) sizes of \textit{PaSST L+M} models with \textit{hop=10 ms}. The base model has \textit{RF=160 ms}, \textit{large RF=640 ms}. \textit{2RF} indicates concatenating the embedding of both receptive fields.}
\label{fig:dm_rf}
\end{figure}

Figure~\ref{fig:dm_rf} compares PaSST L+M models with small \textit{RF=160 ms}, \textit{large RF=640 ms} and concatenation of both (\textit{2RF}). The results show that models with larger RF perform much better on the DCASE task while nearly maintaining their performance on the Maestro task. 

\subsection{Using Mid- and High-level Features}
\begin{figure}[tbh!]
\centering
{\includegraphics[width=1.0\textwidth]{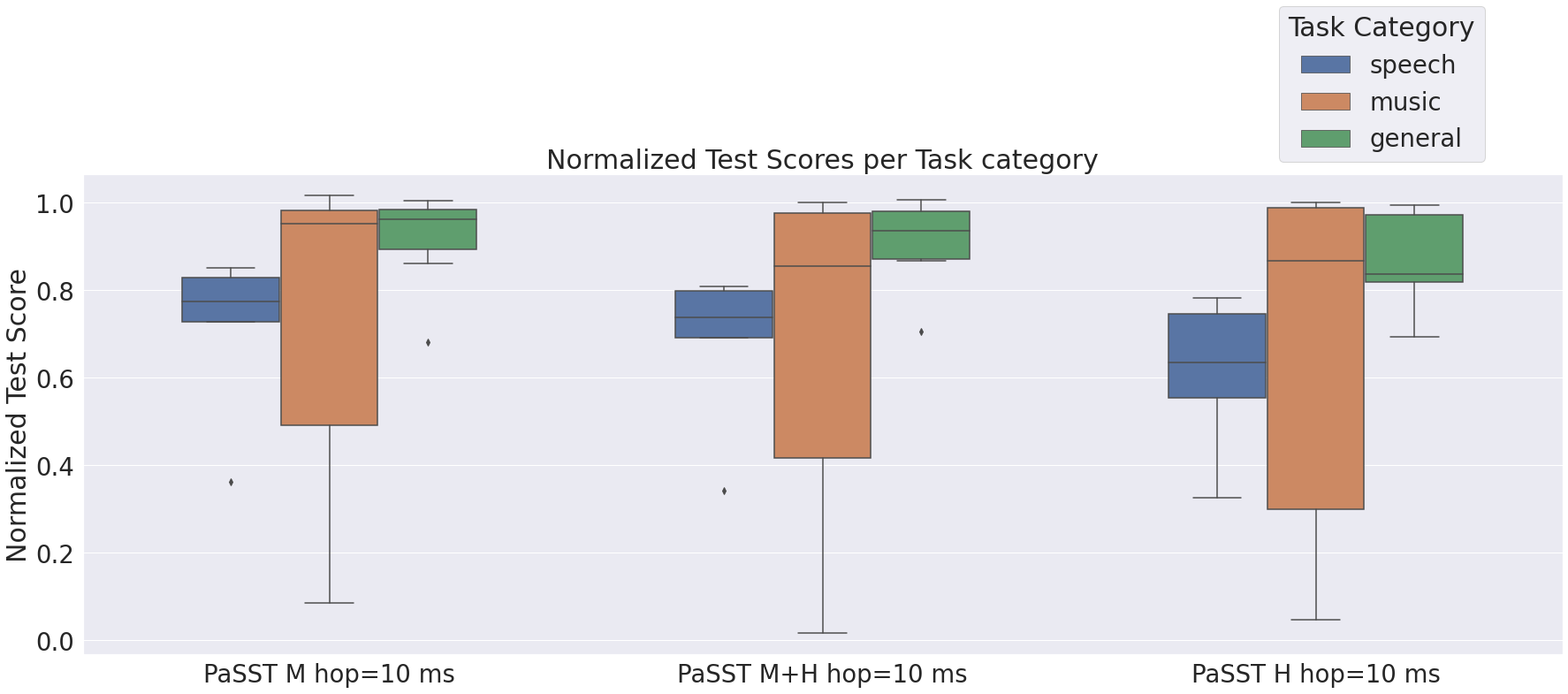}}
\caption{The effect of using different levels of PaSST representations as embeddings. Three different settings are compared: only mid-level M, only high-level H (MLP logits), and  concatenating mid- and high-level representations M+H.}
\label{fig:logits}
\end{figure}

Figure~\ref{fig:logits} compares using medium- and high-level features as embeddings. Using only mid-level features produces the best results across all categories; concatenating mid- and high-level features yields similar results; and using only high-level features yields the worst results.
This suggests that the high-level features of the output PaSST's MLP classifier (see Section~\ref{sec:extract:embed:level}) generalize less because they contain very specific information about Audioset classes. This is demonstrated by the greatest drop in performance in speech tasks that are the furthest away from Audioset classes, as well as music tasks that are further away than the general category.
The mid-level features provide a more abstract representation of the audio, which is beneficial for downstream tasks.

\subsection{Using Finer Temporal Resolution}
\begin{figure}[h!]
\centering
{\includegraphics[width=1.0\textwidth]{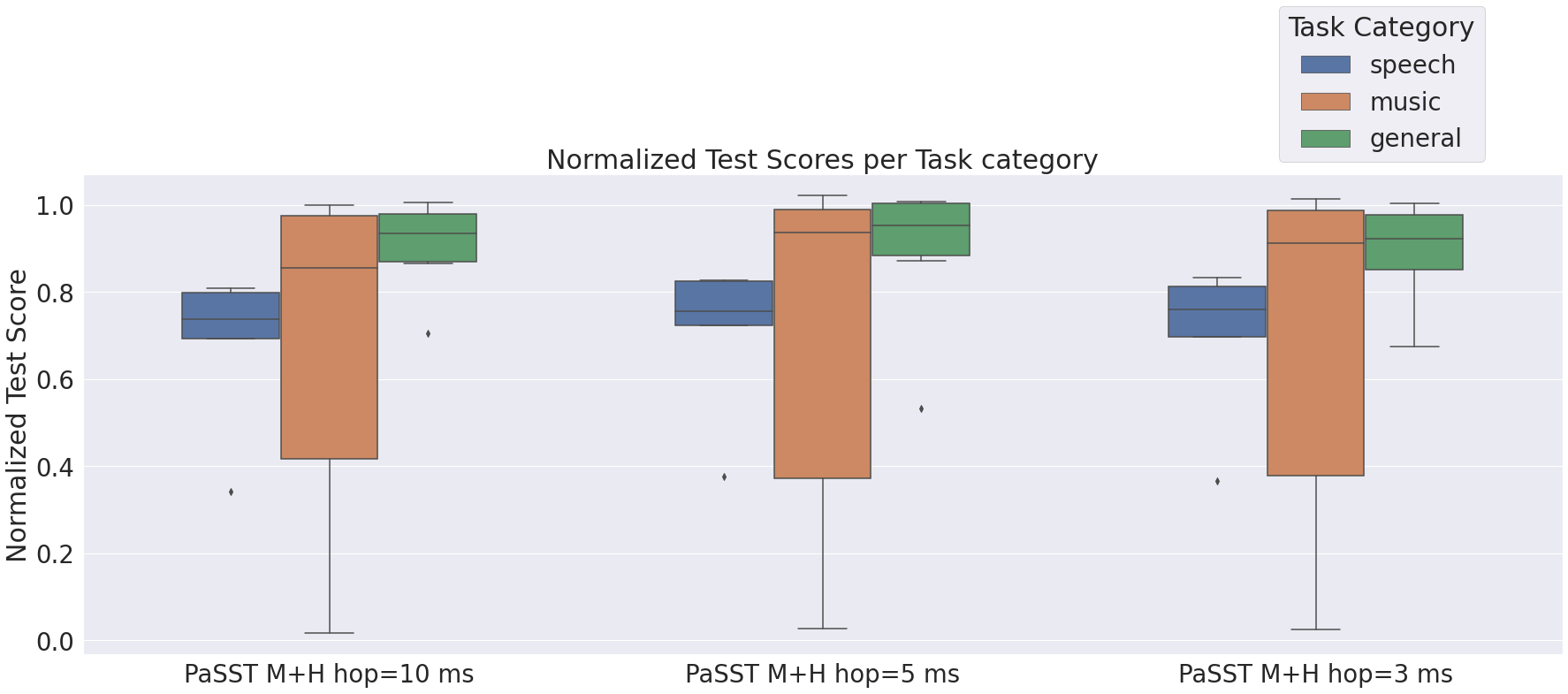}}
\caption{The impact of varying the temporal resolution by changing the hop-size of the STFT window on PaSST performance. 
}
\label{fig:temp_res}
\end{figure}


\begin{table}[!htb]
\centering
\begin{tabular}{cccccc}
\toprule
STFT hop & Seq. / P   &  Speed / P & P Speedup & Infer. &  Audioset \\\hline
    10ms &      1190 /   474  &  \ 23.1 / 88.7    &$3.8\times$& 62.5 & 47.6 \\\midrule
    5ms   &     2388 / 594&  \ \ 6.8 / 62.4   &$9.2\times$& 18.9 &   47.3 \\ \midrule     
    3ms   &     3828 / 1014    & \ \ \ \    3 / 29.7   &$9.9\times$& \ 8.5 &  47.3 \\\bottomrule             
\end{tabular}
\caption{Changing the temporal resolution of PaSST. Seq.: indicates the input sequence length to attention layers.  P: With Patchout. Speed: 10-second-audio training samples per second on an Nvidia Titan RTX GPU. Infer.: Inference speed in samples per second. Audioset: Supervised performance on Audioset in mean Average Precision (mAP).  }
\label{tab:time:resolution}
\end{table}

We trained variants with a finer temporal resolution by decreasing the STFT hop size from \textit{10 ms} to \textit{5 ms} and \textit{3 ms} (as explained in Section~\ref{sec:extract:temp:resolution}). As can be seen in Figure~\ref{fig:temp_res}, the model with a hop size of \textit{5 ms} improves the results for the categories general and speech, and marginally for the category music. Reducing the hop size even further to \textit{3 ms} does not result in additional improvements for any category, but instead worsens the results. Overall, environmental sounds and especially speech benefit from a higher temporal resolution compared to the base PaSST model.  However, this higher resolution comes with a large complexity increase, as shown in Table~\ref{tab:time:resolution}.

\subsection{Single Model Analysis}
\begin{figure}[thb!]
\centering
{\includegraphics[width=1.0\textwidth]{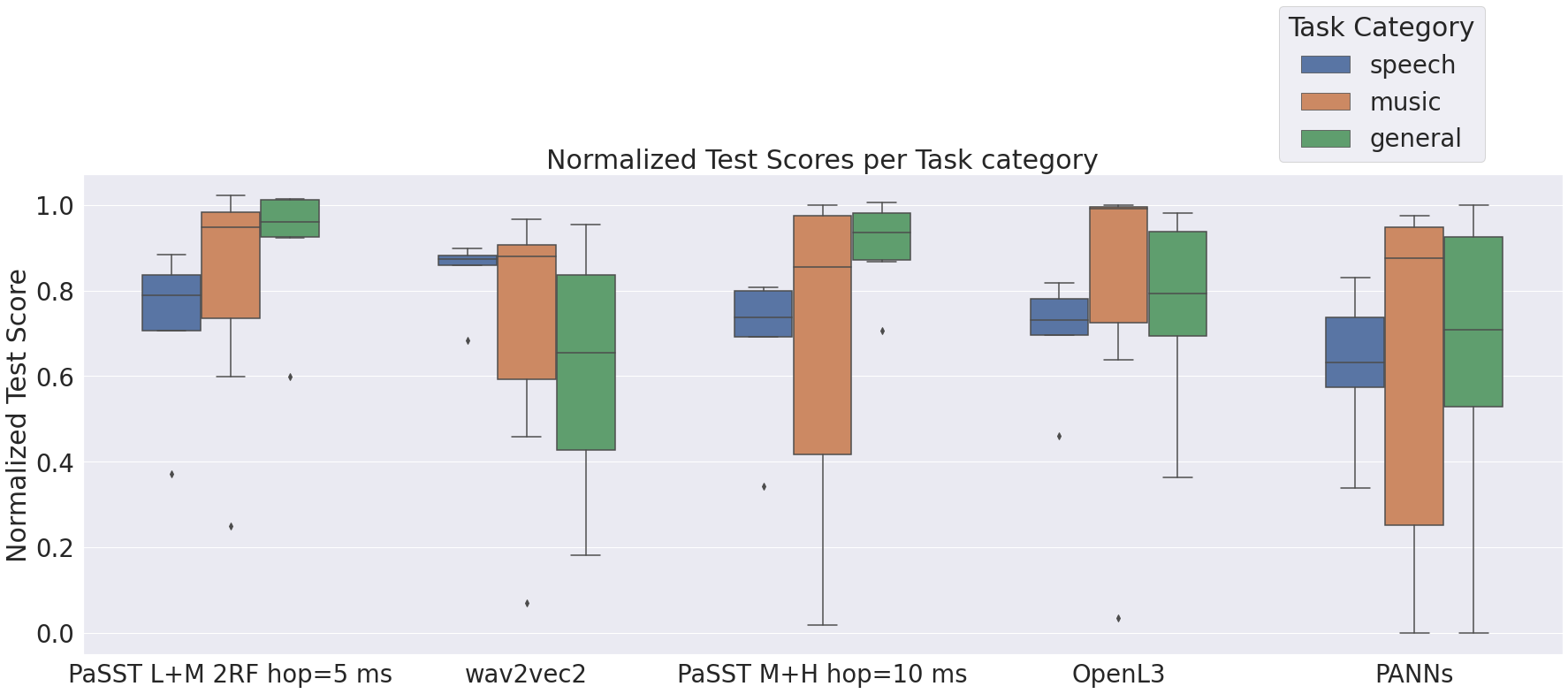}}
\caption{The figure shows a comparison of the models OpenL3 \citep{cramerWSB19openl3}, Wav2Vec2 \citep{baevski2020wav2vec}, PANNs \citep{KongCIWWP20panns}, PaSST~\citep{koutini21passt} in its baseline setting and the improved PaSST model. The y-axis depicts normalized test scores per task. Each model is evaluated by three boxplots for the task categories general, speech, and music.}
\label{fig:category_analysis}
\end{figure}

Figure~\ref{fig:category_analysis} compares different single models on the HEAR categories (see Table~\ref{table:task_catgory}). OpenL3 \citep{cramerWSB19openl3} and PANNs~\citep{KongCIWWP20panns} are based on CNN architectures. PaSST is a transformer model.  Wav2Vec2 \citep{baevski2020wav2vec} is a 1D-CNN with a positional transformer.
PANNs and PaSST are trained on Audioset. OpenL3 is trained on both the audio and video of Audioset in a self-supervised way. Wav2Vec2 is trained in self-supervised fashion on 100K hours of speech from VoxPopuli~\citep{wang2021voxpopuli}. On the speech tasks, Figure~\ref{fig:category_analysis} shows that the specialized model Wav2Vec2 outperforms the models trained on Audioset, PaSST base scores better on average than OpenL3 and PANNs. On music tasks, the distribution of scores has a large variance. OpenL3 shows the best performance. In the general category, PaSST base results in the best performance scoring between 90\% and 100\% of the best scoring models in the challenge in each task.  Comparing CNNs (PANNS) and transformers (PaSST) trained on the same dataset, Figure~\ref{fig:category_analysis} shows that the embeddings from PaSST score better in all categories.
Figure~\ref{fig:category_analysis} additionally shows an improved version of PaSST with the hyper-parameter settings that turned out to be beneficial in the preceding sections. The improved PaSST version (L+M 2RF hop=5ms) scores better in all categories compared to its baseline version. In particular, scores in the music category are improved and have less variance.

\section{Conclusion}

In this paper, we examine different approaches to extract general audio representations using the PaSST transformer model, which was pre-trained on Audioset.
Compared to CNNs trained on the same dataset, our results show that representations extracted by the transformers are more general and capture information that can be better transferred to other tasks.
Despite the variety of Audoiset classes, the results show that mid-level features are more general than high-level logits and can better transfer to downstream tasks.
Finally, we  highlight the trade-off between performance and computational complexity that arises when the temporal resolution is changed.

\section*{ACKNOWLEDGMENT}

The LIT AI Lab is supported by the Federal State of Upper Austria. Gerhard Widmer's work is supported by the European Research Council (ERC) under the European Union's Horizon 2020 research and innovation programme, grant agreement No 101019375 (Whither Music?).

\bibliography{refs}

\appendix

\section{Training Setup}
\label{app:setup}
We use the same download and pre-processing of YouTube Audioset videos as PANNs~\citep{KongCIWWP20panns}, resulting in approximately 2 million 10-second mono audio clips with a sampling rate of 32 Khz. Therefore, we use resampled version to 32Khz of the HEAR 21 datasets~\citep{turian2022hear}.
The default Mel feature parameters are similar to those of AST; we use  a $25$ ms window for Short-Time Fourier Transform (STFT) with a hop length of $10$ ms, resulting in $128$ mel bands and a $1000$ time frames. As explained in Section~\ref{sec:extract:temp:resolution}, we also look into finer time resolution using shorter hop lengths.
We balance the training set as explained in~\citet{koutini21passt}. The reported results on Audioset are evaluated on $18,951$ audio clips. 
We use the AdamW~\citep{LoshchilovH19adamw} optimizer with a weight decay of $10^{-4}$, with a learning rate of $10^{-5}$ and a scheduler as explained~\citet{koutini21passt}.

All the used models are based on DeiT B↑384~\citep{TouvronCDMSJ21deit} pretrained on ImageNet~\citep{dengImagenetLargescaleHierarchical2009} as proposed by \citet{gong21ast}. 
For models with an STFT window hop of 10 ms, we use structured Patchout (see Section~\ref{sec:patchout}) to remove 4 rows (the frequency dimension) and 40 columns (the time dimension). For the models with hops of 5ms and 3ms, we remove 6 rows and 100, 150 columns, respectively. As a result, the sequence lengths and training speedups shown in table~\ref{tab:time:resolution} were achieved.

As explained in \citet{koutini21passt}, we use a number of data augmentations, including Mix-up~\citep{zhangMixupEmpiricalRisk2017}, SpecAugment~\citep{ParkCZCZCL19Specaugment}, waveform rolling, and random Gain

\section{Extended Model Comparison}
\label{app:further:results}

\begin{figure}[tb!]
\centering
{\includegraphics[width=1.0\textwidth]{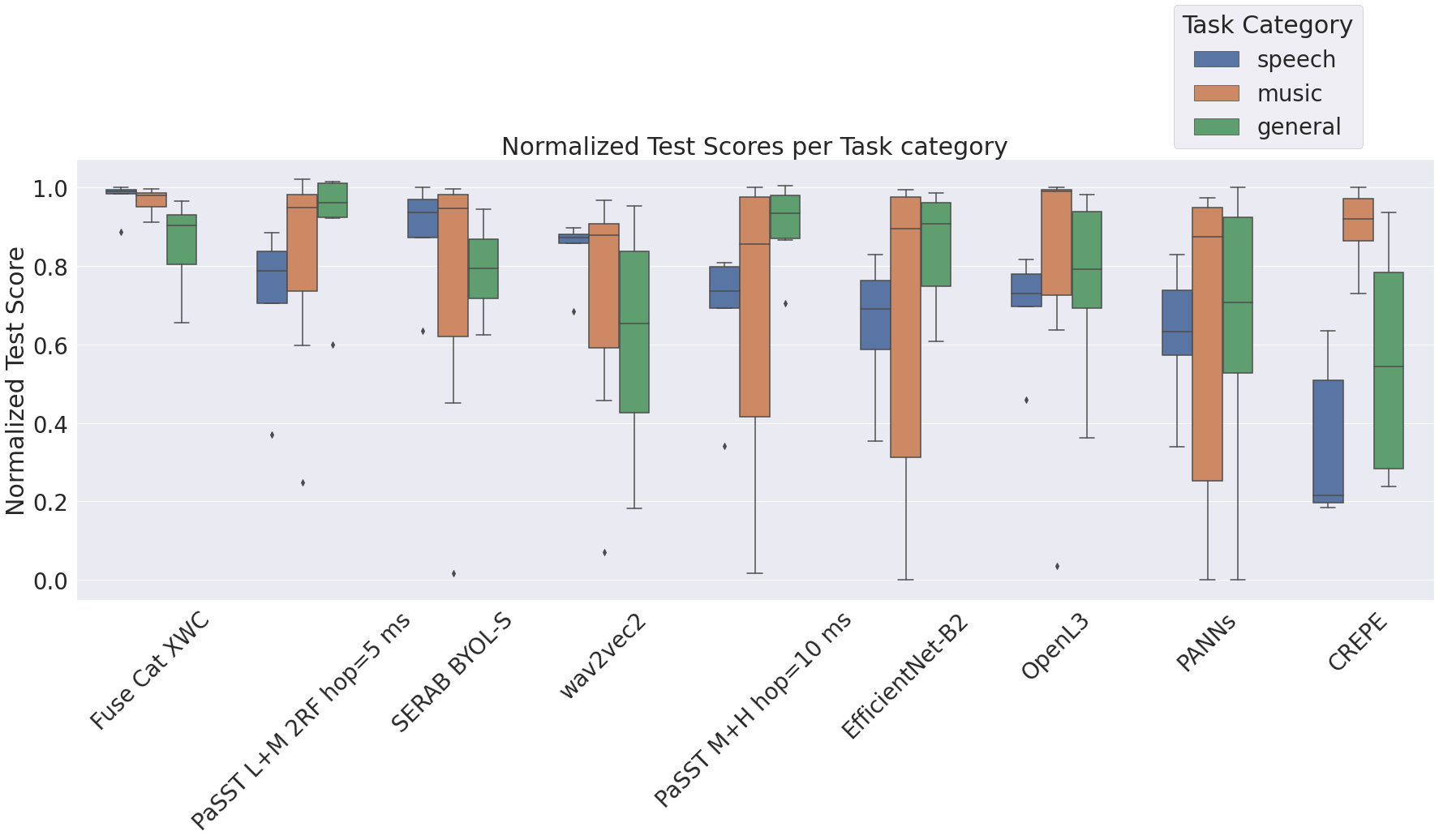}}
\caption{Comparing models that participated in the HEAR 2021 challenge with \textit{PaSST L+M 2RF hop=5 ms}, which is a PaSST model equipped with better hyper-parameter settings for representation extraction (using the results from Section~\ref{sec:results}). }
\label{fig:model_comparison:hear}
\end{figure}

\begin{figure}[tb!]
\centering
{\includegraphics[width=1.0\textwidth]{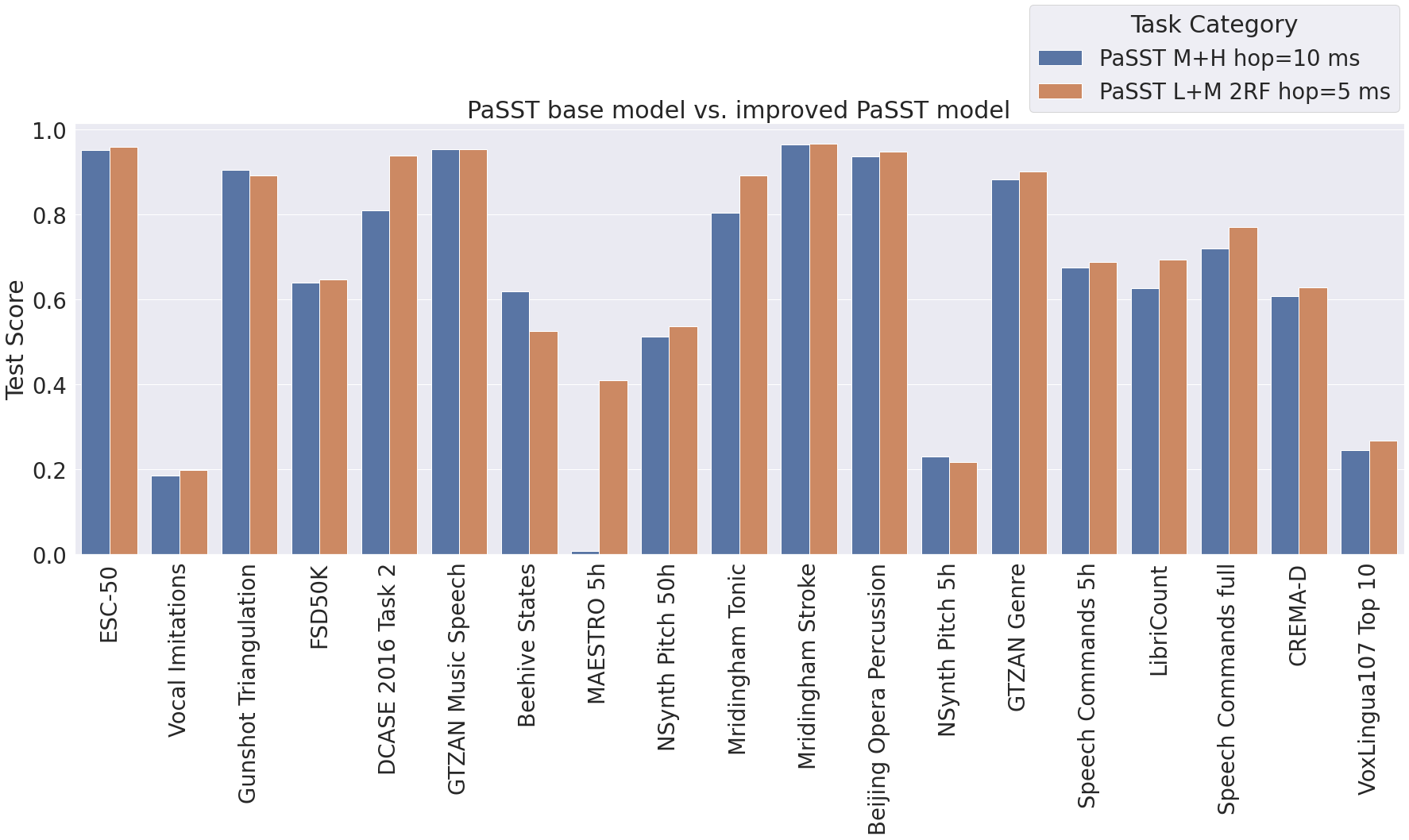}}
\caption{Comparing the results of the PaSST base model (submitted to the HEAR challenge) with the variant equipped with better hyper-parameter settings for representation extraction (using the results from Section~\ref{sec:results}). 
}
\label{fig:base:vs:improved}
\end{figure}

Figure~\ref{fig:model_comparison:hear} compares different models that participated in the HEAR 2021 challenge \citep{turian2022hear}. The models are sorted according to the median normalized test score across all tasks. With respect to this sorting \textit{Fuse Cat XWC}, which is a combination of the three models \textit{HuBERT} \citep{hsu2021hubert}, \textit{wav2vec2} \citep{baevski2020wav2vec} and \textit{CREPE} \citep{kim2018crepe}.

Figure~\ref{fig:base:vs:improved}  highlights the performance gains resulting from using better hyper-parameter settings based on the results presented in Section~\ref{sec:results}. The figure compares the PaSST base model as submitted to the HEAR challenge \emph{PaSST M+H hop=10ms}, with the improved version \emph{PaSST L+M 2RF hop=5ms} on all the tasks.

\end{document}